\documentclass{article}
\usepackage[preprint]{spconf,amsmath,graphicx}

\usepackage{subcaption}
\usepackage{multirow}
\usepackage{booktabs}
\usepackage{url}
\usepackage{cite}

\title{Two-Pass End-to-End ASR Model Compression}
\nameFirstLine{{Nauman Dawalatabad$^*{^\dagger}$, Tushar Vatsal$^*{^\ddagger}$, Ashutosh Gupta$^{\dagger}$,}  
\nameSecondLine{Sungsoo Kim$^\dagger$,  Shatrughan Singh$^\dagger$, Dhananjaya Gowda$^\dagger$, Chanwoo Kim$^\dagger$}%
\thanks{*Equal contribution. $^\ddagger$Work done while at Samsung Research.}%
}
\address{$^\dagger$Samsung Research\\ $^\ddagger$Amazon}
\copyrightnotice{\small{\copyright\ IEEE 2021}}
\toappear{\small{To appear in {\it IEEE ASRU, Dec 13-17, 2021, Cartagena, Colombia.}}}

\begin{document}
%
\maketitle{}
\begin{abstract}

Speech recognition on smart devices is challenging owing to the small memory footprint. Hence small size ASR models are desirable. With the use of popular transducer-based models, it has become possible to practically deploy streaming speech recognition models on small devices~\cite{he2019streaming}. Recently, the two-pass model~\cite{sainath2019two} combining RNN-T and LAS modules has shown exceptional performance for streaming on-device speech recognition. 

In this work, we propose a simple and effective approach to reduce the size of the two-pass model for memory-constrained devices. We employ a popular knowledge distillation approach in three stages using the Teacher-Student training technique. In the first stage, we use a trained RNN-T model as a teacher model and perform knowledge distillation to train the student RNN-T model. The second stage uses the shared encoder and trains a LAS rescorer for student model using the trained RNN-T+LAS teacher model. Finally, we perform deep-finetuning for the student model with a shared RNN-T encoder, RNN-T decoder, and LAS rescorer.  Our experimental results on standard LibriSpeech dataset show that our system can achieve a high compression rate of 55\% without significant degradation in the WER compared to the two-pass teacher model.

\end{abstract}


\section{Introduction}

Nowadays, various smart devices such as smart speakers, mobile phones, health care devices, surveillance devices, smartwatches, and many more have become a part of daily life. Since most of these devices are small in size, it is difficult to have a reasonable-sized keyboard. Moreover, speech is a more natural way to interact with these devices. Hence there is a growing need to deploy speech recognition models on small devices. Further, it is desirable to have streaming output (or online output) to improve user's experience. Recently, there has been tremendous research interest in the field of speech recognition for on-device online streaming scenarios \cite{he2019streaming,sainath2019two,ganesh_ondevice,samsung_streaming, garg2019asru, infra_paper, sr_vd_is2020}.

Most smart devices have a low memory footprint. Deploying large size Automatic Speech Recognition (ASR) models is infeasible on such devices. Hence, many applications transfer the user's speech data to the large server, processes the spoken sentence, and then revert back with the proper response. However, this protocol incurs high latency due to data transfer time between the device and the server. Furthermore, transferring user's data online may create data privacy concerns. Hence it is desirable to have the ASR model operating on-device.

End-to-End (E2E) approaches have achieved comparable performances to the traditional systems by having a single E2E model instead of separate components~\cite{kim2020review_ondevice}. These approaches includes Connectionist Temporal Classification (CTC)~\cite{graves2006connectionist}, Attention-based Encoder-Decoder (AED) \cite{chan2015listen}, and Recurrent Neural Network Transducer (RNN-T)~\cite{he2019streaming} based models. The AED models need the complete audio beforehand to start decoding the audio, making AED models unsuitable for streaming purposes. The RNN-T and CTC models can be deployed for streaming scenarios due to their frame-synchronous output nature. RNN-T model is preferred over CTC because it has token-dependency modeling with a prediction network. Additionally, CTC also has a strong assumption of output tokens to be independent, which makes RNN-T more preferable over CTC networks. However, RNN-T models consume significant memory as it contains transcription, prediction, joint networks and have higher decoding latency because of framewise predictions.

Online speech recognition task demands both smaller model size and lower latency for small memory footprint devices, necessitating the need for model compression techniques for such models to be deployable for streaming purposes.
Knowledge distillation ~\cite{Gou_2021,chebotar2016distilling,hinton2015distilling} is a popular approach for compression of neural models. It refers to the idea of model compression by training a smaller model (student), step-by-step, with the help of an already trained bigger model (teacher). The student model can be trained to replicate the outputs of the teacher model at different levels. 
The student model with fewer parameters tries to mimic the teacher model's behavior (with a large number of parameters). 

Knowledge distillation (KD)  has been successfully employed in many speech recognition tasks.
Authors in~\cite{fukuda2017efficient} improves performance of a Convolutional Neural Network (CNN)~\cite{albawi2017understanding} with the help of ensemble of VGG~\cite{simonyan2014very} and LSTM \cite{sak2014long, sherstinsky2020fundamentals} as the teacher models.
KD has also been helpful in attention-based models for Neural Machine Translation (NMT) task~\cite{kim2016sequence} and maximum mutual information-based training~\cite{wong2016sequence}.
Recently, authors in~\cite{inaguma2021source} also use KD for E2E speech translation task.
In~\cite{kim2017improved}, KD is used for distilling knowledge from the bidirectional to unidirectional CTC models. Authors in~\cite{takashima2018investigation} compare the performance of sequence-level and frame-level KD losses for CTC models.  In \cite{kurata2020knowledge}, an efficient strategy to transfer knowledge from the offline RNN-T model to streaming one is proposed. Recently, authors in~\cite{panchapagesan2020efficient} proposes a simple and efficient approach for KD with RNN-T models, which requires less memory overhead during training, leading to faster convergence. 

In this paper, we propose a three-stage approach for compressing a two-pass streaming ASR model using KD.
In the first stage, we perform KD to train a student RNN-T model using a trained teacher RNN-T model~\cite{panchapagesan2020efficient}. 
The second stage trains an additional Listen, Attend, and Spell (LAS)~\cite{las} module with an RNN-T encoder.
Finally, in the third stage, we finetune the model by combining the RNN-T encoder, the RNN-T decoder, and the LAS module.
With the proposed approach, we achieve 55\% of compression without significant degradation in performance.
We also conduct various experimental analyses to study the effect of KD in different stages of the proposed approach.

\section{Two-pass E2E Model}
\label{sec:background}
The two-pass model~\cite{sainath2019two} consists of the RNN-T and the LAS modules.
Each of the modules is described in the following sections.

\subsection{Recurrent Neural Network Tranducer }
\label{ssec:rnnt}
The RNN-T is an encoder-decoder sequence-to-sequence prediction model that is widely used for end-to-end streaming speech recognition systems~\cite{graves2012sequence}. The RNN-T model comprises the transcription, predication, and a joint network.
The RNN-T training aims at minimizing the negative log posterior probability as given by the following equation:
\begin{equation}
L_{RNN-T} = -\ln~P(\boldsymbol{y^{*}}|\boldsymbol{x})
\end{equation}
where $\boldsymbol{y^{*}}$ is a target sequence and $\boldsymbol{x}$ is input sequence. The probability $P(\boldsymbol{y^{*}}|\boldsymbol{x})$ is computed using $T \times U$  probability lattice, where $T$ denotes the input sequence length and $U$ is the output sequence length~\cite{graves2012sequence}. Each node in the lattice outputs a probability distribution $P(k|t,u)$ over target vocab including a blank symbol $\phi$, where $k$ represents the output label. 
The RNN-T model can be represented with the following set of equations:
\begin{equation}
\begin{gathered}
    h_{t} = LSTMs(x_{t}, h_{t-1})\\
    g_{u} = LSTMs(y_{u}, g_{u-1})\\
    z_{t,u} = V^{T}\tanh(Wh_{t} + Ug_{u})\\
    P(k|t,u) = softmax(z_{t,u}) \label{RNNTeq}
\end{gathered}
\end{equation}
Equation~\eqref{RNNTeq} correspond to the outputs from the transcription, prediction and joint networks (in order). The $P(\boldsymbol{y}|\boldsymbol{x})$ is calculated by summing over all possible alignments $\boldsymbol{a}$.
\begin{equation}
    P(\boldsymbol{y}|\boldsymbol{x}) = \sum_{\boldsymbol{a} \in B^{-1}(y)} P(\boldsymbol{a}|\boldsymbol{x})
\end{equation}
where $B$ is a function that removes blank symbols from alignments. Summing over all possible alignments is intractable, and hence forward-backward algorithm is used to compute the same. More information on the RNN-T model can be found in~\cite{graves2012sequence}.

\begin{figure}[t]
    \centering
    \includegraphics[scale=0.3]{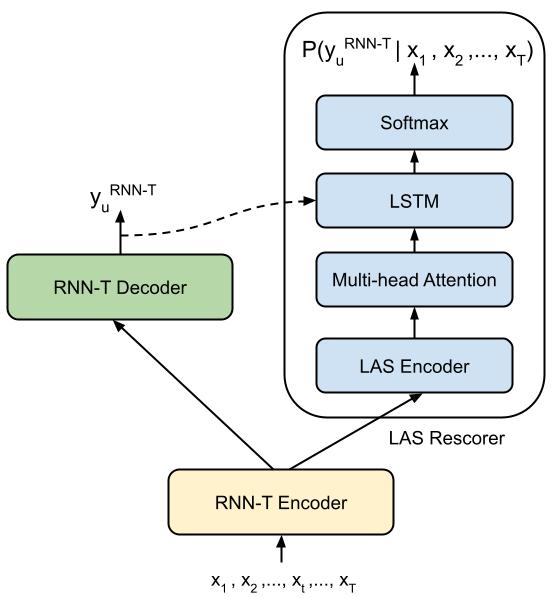}
    \caption{Architecture of Two-pass model.}
    \label{fig:2pass}
\end{figure}

\begin{figure*}[t]
  \centering
  \begin{subfigure}[]{0.33\textwidth}
    \centering
    \includegraphics[width=0.852\linewidth]{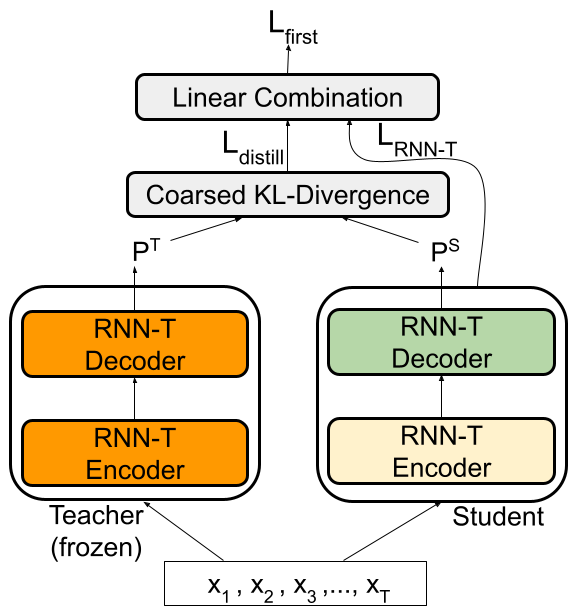}
    \caption{Stage-1: Teacher frozen}
    \label{fig:kd-stage-1}
  \end{subfigure}%
  \begin{subfigure}[]{0.33\textwidth}
    \centering
    \includegraphics[width=0.833\linewidth]{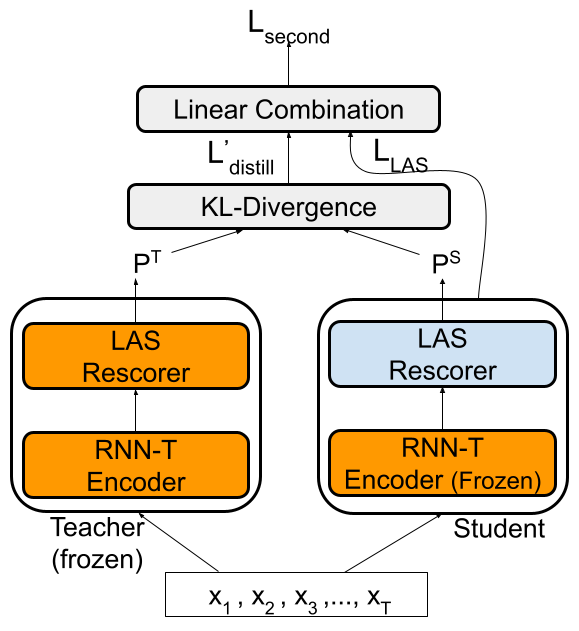}
    \caption{Stage-2:  Student (encoder) frozen}
    \label{fig:kd-stage-2}
  \end{subfigure}
  \begin{subfigure}[]{0.33\textwidth}
    \centering
    \includegraphics[width=0.9\linewidth]{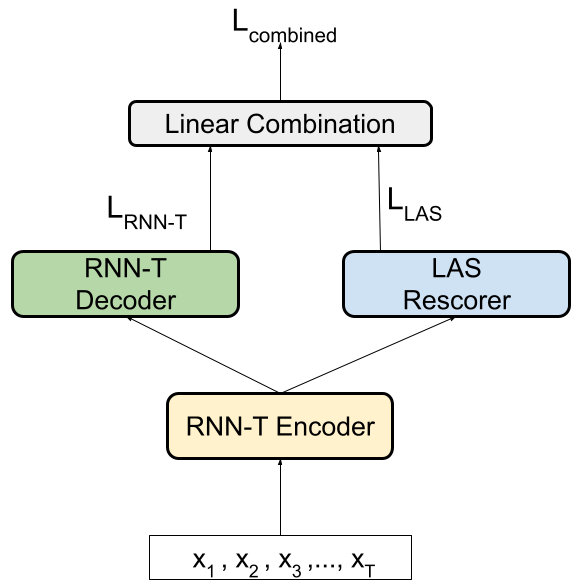}
    \caption{Stage-3: Deep fine-tuning}
    \label{fig:kd-stage-3}
  \end{subfigure}
  \caption{The proposed three-stage approach for compressing the two-pass ASR model. Blocks colored in orange are kept frozen during training. (a) Stage-1: KD teacher-student training is used to trained the student RNN-T model. Coarsed KL-Divergence represents distillation loss given in Equation~\eqref{firsttoteq}. (b) Stage-2: An LAS rescorer is added to the RNN-T encoder and trained using KD. (c) Stage-3: The two-pass student model is deep-finetuned.}
  \label{fig:KD_production}
\end{figure*}

\subsection{Two-pass model with RNN-T and LAS}
\label{sec:twopass}

As shown in Fig.~\ref{fig:2pass}, the two-pass ASR model consists of an additional encoder layer on top of the shared RNN-T encoder. It also has a LAS decoder in addition to RNN-T model components. The LAS module is used to improve the streaming ASR output from RNN-T. Training two-pass model~\cite{sainath2019two} is not trivial because of its known issue of non-converging loss.  The training of the two-pass model is carried out in three stages~\cite{sainath2019two} as follows:
\begin{enumerate}
    \item Train RNN-T model (Section~\ref{ssec:rnnt})~\cite{graves2012sequence}.
    \item Train additional two-pass components with cross-entropy loss using pre-trained shared RNN-T encoder, which remains frozen.
    \item Deep-finetune the entire network using RNN-T and LAS loss.
\end{enumerate}

\section{Proposed Two-pass model Compression}
\label{sec:proposed}

Knowledge distillation is used for compression by using a teacher model, where another student model, smaller in size, is trained to mimic the teacher model. To mimic the teacher model, KL-Divergence between the probability distribution of teacher and student model is used as distillation loss which is combined with original training loss using some trade-off. Similar to training two-pass model, KD with two-pass models is also not trivial. Motivated from the training procedure of the two-pass model, the following is the proposed three-stage approach used for compression of the two-pass model as shown in Fig.~\ref{fig:KD_production}.

\subsection{Stage-1: KD training}

In the first stage, we train a student RNN-T model using KD from the trained RNN-T teacher model, which is kept frozen as shown in Fig.~\ref{fig:kd-stage-1}. 
As discussed in Section~\ref{ssec:rnnt}, the RNN-T loss is computed by summing over all possible alignments. Thus KL-Divergence between probability distributions of teacher and student model can be given by:
\begin{equation}
    L_{distill} = \sum_{t,u}\sum_{k} P^{T}(k|t,u)\ln\frac{P^{T}(k|t,u)}{P^{S}(k|t,u)}
\end{equation}
where $P^{T}$ and $P^{S}$ are probability distributions of teacher and student RNN-T models at time step $t$. As we can see, distillation loss computation requires an additional memory size of $\mathcal{O}(T \times U \times K)$ per utterance which is too expensive and impractical. Hence the  $K$ dimensional probability distribution is reduced to three dimensions by only taking in account conditional probabilities of $y_{u+1}$, $\phi$ and remainder tokens at any lattice node $(t,u)$. The corresponding transition probabilities are as follows~\cite{panchapagesan2020efficient}:
\begin{equation}
    \begin{gathered}
        P_{y}(t,u) = P(y_{u+1}|t,u) = y(t,u)\\
        P_{\phi}(t,u) = P(\phi|t,u) = \phi(t,u)\\
        P_{r}(t,u) = 1 - P_{y}(t,u) - P_{\phi}(t,u) \label{probseq}
    \end{gathered}
\end{equation}
Thus distillation loss can be defined as KL-Divergence between probability distributions of teacher and student model as follows:
\begin{equation}
    L_{distill} = \sum_{t,u}\sum_{l \in\{y,\phi,r\}}P^{T}_{l}(t,u)\ln\frac{{P^{T}_{l}}(t,u)}{P^{S}_{l}(t,u)}
\end{equation}
The overall loss in the Stage-1 of the knowledge distillation training is given by a linear combination of $L_{RNN-T}$ and $L_{distill}$ as follows:
\begin{equation}
    L_{first} = \beta L_{distill} + (1 - \beta) L_{RNN-T} \label{firsttoteq}
\end{equation}
where $\beta$ represents the distillation loss weight.

\subsection{Stage-2: KD training}
\label{ssec:kd-stage-2}
As shown in Fig.~\ref{fig:kd-stage-2}, in Stage-2 KD training, a LAS rescorer module is added to the RNN-T encoder obtained from Stage-1 training. 
The RNN-T encoder of the student model is kept frozen, and only the parameters of LAS rescorer modules are updated during backpropagation. The teacher model is kept frozen during KD training.

The distillation loss in the Stage-2 KD training is given by  KL-Divergence between two-pass LAS rescorer's probability distribution of teacher and student models as follows:
\begin{equation}
        L_{distill}^{'} = \sum_{t}\sum_{k} P^{T}_{2}(k|t)\ln\frac{P^{T}_{2}(k|t)}{P^{S}_{2}(k|t)} 
\end{equation}
where $P^{T}_{2}$ and $P^{S}_{2}$ are second pass LAS rescorer's probability distributions of teacher and student model respectively at each time step $t$ over target vocab $k$ including a blank symbol. The total loss for the Stage-2 training is  given by
\begin{equation}
    L_{second} = \gamma L_{distill}^{'} + (1 - \gamma) L_{LAS} \label{eq:2ndstageloss}
\end{equation}
where $\gamma$ is a tunable hyperparameter which manages trade-off between these two losses.

\subsection{Stage-3: Deep finetuning}

As shown in Fig.~\ref{fig:kd-stage-3}, in the third stage, we combine the RNN-T encoder (shared), the RNN-T decoder, and the LAS rescorer. 
Deep-finetuning is performed on this model. 
The combined loss for training used in the Stage-3 is given by
\begin{equation}
    L_{combined} = \lambda L_{RNN-T} + (1 - \lambda) L_{LAS} \label{comblosseq}
\end{equation}
where $\lambda$ is a tunable hyperparameter which is set to 0.5 in our trainings throughout this work.

Evaluation can be done via ``$2^{nd}$ beam search'' mode or ``rescoring'' mode~\cite{sainath2019two}. In this paper, we use the ``rescoring'' mode for evaluation. In rescoring mode, we run LAS decoder on the top-K hypotheses predicted by the RNN-T decoder in teacher-forcing mode~\cite{sainath2019two} with attention on additional encoder layer output to rescore the beams. We then pick the beam with the highest score from the LAS module. From this point onwards, we will use the term LAS rescorer for the additional second pass module, comprising of an additional LSTM layer followed by LAS decoder, as also shown in  Fig.~\ref{fig:2pass}.

\section {Experimental Setup}
\subsection{Dataset, features, and augmentation}
We train our models on the standard LibriSpeech dataset~\cite{panayotov2015librispeech} containing 960 hours of training speech data and evaluate on test-clean and test-other datasets containing 2620 and 2939 utterances, respectively. We use Tensorflow 2 Keras API to build the complete training pipeline. The target vocab contains around 10K Byte Pair Encoding (BPE)~\cite{sennrich2015neural} units and are embedded in 621-dimensional space for the teacher model and 390-dimensional space for the smaller model. We use 40 dimensional Mel Frequency Cepstral Coefficients (MFCCs)~\cite{wanli2013research} computed from a 25~msec window and shifted every 10~msec as an input to the encoder network. We use a popular SpecAugment~\cite{park2019specaugment} technique for data augmentation during training.
The beam size of 8 is used for decoding. For two-pass models, evaluation is done in rescoring mode.

\subsection{RNN-T model} 
\label{rnntdetails:sec}
As discussed in Section~\ref{ssec:rnnt}, the RNN-T model has three components, transcription network, prediction network, and joint network. It is an encoder-decoder framework where the transcription network acts as an encoder, and the prediction network coupled with the joint network acts as a decoder. The transcription network consists of 6 unidirectional LSTM layers. We use max-pooling of 2 in the first three layers, with an overall factor of 8, to reduce the training time and latency of the model. We use dropout~\cite{srivastava2014dropout} of 0.2 after each layer except the first one. To ensure proper convergence, we gradually add layers in the initial stages of training starting from 2 layers~\cite{zeyer2018improved}. The prediction network comprises an embedding layer followed by one LSTM layer, which acts as a conditional Language Model (LM) model. The joint network consists of two dense layers, one each for transcription and prediction output, respectively. It also consists of an additional dense layer to project the features into vocab space which acts as logits for RNN-T loss calculation. The initial learning rate of $5\times10^{-4}$ is used with Adam optimizer~\cite{kingma2014adam}, which is decayed exponentially by a factor of $0.9$ after every 20K steps. Student RNN-T model is trained with the same hyperparameters with an additional distillation loss linearly combined with the RNN-T loss as given in Equation~\eqref{firsttoteq}.

\vspace{-1em}
\subsection{Two-pass model}
The two-pass model contains an additional LAS rescorer and is trained in three stages as discussed in Section~\ref{sec:twopass}. The first stage model is trained with architecture details similar to Section~\ref{rnntdetails:sec}. In the second stage, the shared encoder, i.e., the transcription network of the RNN-T model is frozen and is used to train the additional encoder and LAS rescorer as mentioned in Section~\ref{ssec:kd-stage-2}. The additional encoder layer consists of one unidirectional LSTM layer. The LAS rescorer consists of an embedding layer and one unidirectional LSTM layer. It uses a 4-headed dot attention layer~\cite{vaswani2017attention}. The learning rate strategy used to train the second stage is similar to the RNN-T model.

\begin{table*}[t]
\caption{Notations used for each of the models are mentioned. The number of parameters in each of the modules for different models and the percentage reduction in the number of parameters with respect to the Teacher model are also mentioned. The subscript $i$ denotes the stage number in knowledge distillation training.}
\centering
\resizebox{0.87\textwidth}{!}{%
\begin{tabular}{c|c| c c c c }
\toprule
\begin{tabular}[c]{@{}c@{}}\textbf{Model}\\ \textbf{Description}\end{tabular}                                     & \textbf{Model-ID}       & \begin{tabular}[c]{@{}c@{}}\textbf{\# params in first pass} \\ \textbf{(RNN-T) module}\end{tabular} & \begin{tabular}[c]{@{}c@{}}\textbf{\# params in second pass} \\ \textbf{(additional) module}\end{tabular} & \textbf{Tot. \# params} & \begin{tabular}[c]{@{}c@{}}\textbf{Reduced size} \\ \textbf{w.r.t. $T_i$ (\%)}\end{tabular} \\ 
\midrule \midrule
\multirow{2}{*}{\begin{tabular}[c]{@{}c@{}}Large size model\\ (also used as Teacher)\end{tabular}}        & $T_1$          & 72M                                                                               & --                                                                                      & 72M            & --                                                                        \\ 
                                                                                                & $T_2$, $T_3$   & 72M                                                                               & 38M                                                                                     & 110M           & --                                                                        \\ 
\midrule
\multirow{2}{*}{\begin{tabular}[c]{@{}c@{}}Small size\\ Model \end{tabular}}                    & $M_1$          & 32M                                                                               & --                                                                                      & 32M            & 55\%                                                                      \\ 
                                                                                                & $M_2$, $M_3$   & 32M                                                                               & 18M                                                                                     & 50M            & 55\%                                                                      \\ 
                                                                                                \midrule
\multirow{2}{*}{Student Model}                                                                  & $S_1$          & 32M                                                                               & --                                                                                      & 32M            & 55\%                                                                      \\ 
                                                                                                & $S_2$, $S_3$   & 32M                                                                               & 38M                                                                                     & 70M            & 36\%                                                                      \\  
\bottomrule
\end{tabular}}
\label{tab:notation-size}
\end{table*}

\vspace{-1em}
\subsection{Notations used}

We denote the large model, which is also used as the teacher model as $T$. 
We also train a smaller size model compared to $T$, which is represented by $M$.
Note that both $T$ and $M$ models are trained independently without using knowledge distillation, as explained in Section~\ref{sec:background}.
The student model represented by $S$ is trained with the proposed distillation approach as explained in Section~\ref{sec:proposed}. 
We compare the performances of the teacher model ($T$) and the smaller model ($M$), with that of the student model ($S$). 

Table~\ref{tab:notation-size} shows the notations used to denote different models along with the corresponding model sizes in terms of the number of parameters.
The training stage number for different models are mentioned as a subscript.
For the Stage-1 training, the RNN-T teacher model is denoted by $T_{1}$, the smaller RNN-T model is denoted by $M_{1}$, and the student model which is trained from $T_{1}$ with the loss given in Equation~\eqref{firsttoteq}  is denoted by $S_{1}$. For Stage-2 of the proposed approach, the teacher model $T_{2}$ is trained using a pre-trained model $T_{1}$ and the smaller model $M_{2}$ is trained using a pre-trained model $M_{1}$. Similarly, the student model in Stage-2 trained with the loss given in Equation~\eqref{eq:2ndstageloss} is denoted by $S_{2}$. 
In Stage-3 training of the proposed approach, the two-pass models minimizes combined loss given in Equation~\eqref{comblosseq} are denoted by $T_{3}$, $M_{3}$, and $S_{3}$.

\begin{table}[t]
\caption{WERs(\%)/ SERs(\%) on LibriSpeech for teacher model ($T_i$), smaller model ($M_i$), and student model ($S_i$).}
\centering
\resizebox{0.36\textwidth}{!}{%
\begin{tabular}{ c | c c| c c } 
 \toprule
 \multirow{2}{*}{\textbf{Model-ID}} & \multicolumn{2}{|c}{\textbf{test-clean}} & \multicolumn{2}{|c}{\textbf{test-other}} \\\cline{2-5}
      & \textbf{WER} & \textbf{SER} & \textbf{WER} & \textbf{SER} \\ 
 \midrule \midrule
 $T_{1}$ & 6.81 & 59.69 & 18.23 & 83.09 \\
 $M_{1}$ & 7.52 & 61.53 & 20.03 & 84.79 \\
 $S_{1}$ & 6.92 & 59.08 & 18.83 & 82.58 \\
 \midrule
 $T_{2}$ & 5.80 & 53.63 & 16.79 & 79.62 \\
 $M_{2}$ & 6.60 & 57.44 & 18.72 & 83.57 \\
 $S_{2}$ & 6.24 & 55.95 & 17.77 & 81.35 \\
 \midrule
 $T_{3}$ & 5.87 & 53.55 & 16.92 & 80.16 \\
 $M_{3}$ & 6.59 & 57.02 & 18.47 & 82.51 \\
 $S_{3}$ & \textbf{5.81} & \textbf{52.67} & \textbf{17.27} & \textbf{79.72} \\
 \bottomrule
\end{tabular}
}
\label{tab:main-results}
\end{table}

\vspace{-1em}
\section {Results}

In this section, we demonstrate the effectiveness of the proposed compression approach on the LibriSpeech dataset.

\vspace{-0.2cm}
\subsection{Performance of proposed approach}
\label{ssec:performance}
Table~\ref{tab:main-results} reports the performance of different models in terms of Word Error Rate (WER) and Sentence Error Rate (SER) for the test-clean and test-other parts of the LibriSpeech dataset, respectively.
The WER/SER mentioned are slightly higher compared to other offline systems or those using external language model.
We not only report the final performance after completing the three KD stages but also report the performances in the intermediate stages, i.e., after Stage-1 and Stage-2 KD training. This gives a detailed overview of the changes in the performances.

It can be seen from the Table~\ref{tab:main-results} that the smaller models $M_1$, $M_2$, and $M_3$ shows significantly degraded performance compared to models $T_1$, $T_2$, and $T_3$ respectively. 
This clearly shows that directly reducing the number of parameters may have an adverse effect on WER performance.

It can be seen from Table \ref{tab:main-results} that the  $S_{1}$ RNN-T model achieves around 8\% relative improvement in WER over same-sized model $M_{1}$, by using distillation loss with teacher model $T_{1}$ on the test-clean part. 
While going from $T_{1}$ to $M_{1}$, a relative degradation of 10\% in WER  with respect to the model $T_1$ is observed due to 55\% reduced parameters. Since this gap is high, the model $M_1$ is not usable.
After applying knowledge distillation with teacher-student training (using Teacher model $T_1$),  our student model $S_1$ performs comparable to the teacher model $T_1$ with minor degradation in WER while maintaining a high compression rate of 55\% compared to $T_1$. A similar trend can be observed for the test-other part.

In the second stage, it can be seen that directly reducing the number of parameters ($M_2$) results in degradation in performance for test-clean by 13.7\% (relative) while with the proposed KD approach with model $S_2$ this degradation is reduced to 7.5\% relative while maintaining the same compression rate. 
This shows that KD training is also helpful even in the case of an attention-based decoder where the student model has the flexibility to attend to different parts.
A similar trend can also be observed for the test-other part. 
This further confirms that the knowledge distillation indeed helps to transfer the teacher's knowledge. 

In the third stage, we combine different modules to form a two-pass model. 
The two-pass models have degradation of 12.26\% in WER  (relative) when model size is directly reduced to 55\% from $T_{3}$ to $M_{3}$.
With deep-finetuning, the student model $S_3$ shows performance even better than the Teacher model ($T_3$) while having a model size that is 36\% less than the Teacher model ($T_3$). 
This shows that the proposed KD training approach is indeed helpful in compressing the two-pass model without loss in WER performance.

Furthermore, we also conduct an experiment where we first trained the two-pass teacher model completely as given in Section \ref{sec:twopass}. The modules from the fully trained two-pass teacher model (i.e., $T_3$) are taken for performing knowledge distillation training in different stages in the proposed approach. We observe a similar trend in the performance.  

\begin{table}[t]
    \centering
    \caption{WERs(\%) / SERs(\%) for student RNN-T model $S_1$ with different values of $\beta$.}
    \resizebox{0.36\textwidth}{!}{%
    \begin{tabular}{ c|c c|c c } 
     \toprule
     \multirow{2}{*}{\textbf{$\beta$ value}} & \multicolumn{2}{|c}{\textbf{test-clean}} & \multicolumn{2}{|c}{\textbf{test-other}} \\\cline{2-5}
          & \textbf{WER} & \textbf{SER} & \textbf{WER} & \textbf{SER} \\ 
     \midrule \midrule
     $0.0$ & 7.52 & 61.53 & 20.03 & 84.79 \\
     $1\times10^{-4}$ & 7.47 & 61.41 & 20.17 & 84.28 \\ 
     $1\times10^{-3}$ & 7.25 & 60.42 & 19.30 & 84.14 \\
     $1\times10^{-2}$ & \textbf{6.92} & \textbf{59.08} & \textbf{18.83} & \textbf{82.58} \\
     $1\times10^{-1}$ & 7.07 & 60.38 & 18.87 & 83.29 \\
     $1.0$ & 10.02 & 71.56 & 21.92 & 87.96 \\
     \bottomrule
    \end{tabular}
    }
    \label{tab:beta}
\end{table}

\begin{table}[t]
\caption{WERs(\%)/ SERs(\%) for the case when distillation is omitted from the Stage-2 in the proposed approach. The KD is performed in the Stage-2 for the models $S_{2}$ and $S_{3}$, whereas it not performed for the models denoted by $S_{2}^{'}$ and $S_{3}^{'}$.}
\centering
\resizebox{0.36\textwidth}{!}{%
\begin{tabular}{ c|c c| c c } 
 \toprule
 \multirow{2}{*}{\textbf{Model-ID}} & \multicolumn{2}{|c}{\textbf{test-clean}} & \multicolumn{2}{|c}{\textbf{test-other}} \\\cline{2-5}
      & \textbf{WER} & \textbf{SER} & \textbf{WER} & \textbf{SER} \\ 
 \midrule
 \midrule
 $T_{2}$ & 5.80 & 53.63 & 16.79 & 79.62 \\
 ${S}_{2}^{'}$ & 6.19 & 54.81 & 17.82 & 81.18 \\
 $S_{2}$ & 6.24 & 55.95 & 17.77 & 81.35 \\
 \midrule
 $T_{3}$ & 5.87 & 53.55 & 16.92 & 80.16 \\
 $S_{3}^{'}$ & 5.92 & 52.82 & 17.52 & 80.81 \\
 $S_{3}$ & \textbf{5.81} & \textbf{52.67} & \textbf{17.27} & \textbf{79.72} \\
 \bottomrule
\end{tabular}
}
\label{tab:ablation-stage2}
\end{table}

\vspace{-0.3cm}
\subsection{Tuning hyperparameter $\beta$ in Stage-1}
\vspace{-0.1cm}
As discussed in Section~\ref{sec:twopass}, $\beta$ is a hyperparameter used for distillation loss in Stage-1 training of the proposed approach.
We found that this hyperparameter needs to be appropriately tuned to achieve good performance. Hence, we analyzed the performance of model $S_1$ under different values of $\beta$.
The performance on model $S_1$ when trained with different values of $\beta$ is shown in Table~\ref{tab:beta}.
It can be seen that extreme values of $\beta$ result in performance degradation. Optimal performance is seen at $1\times10^{-2}$, which is also used in this work.

\vspace{-0.25cm}
\subsection{Ablation study on knowledge distillation in Stage-2}
We conducted an experiment to study the effect of knowledge distillation in the Stage-2 of the proposed approach.
In this experiment we consider two different variants of student model as follows: (i) Distillation is performed in Stage-2 (denoted by $S_2$ and $S_3$), similar to Section~\ref{ssec:performance}, and (ii) No distillation is performed in Stage-2 (denoted by $S_{2}^{'}$ and $S_{3}^{'}$). 

The results of this experiments are shown in Table \ref{tab:ablation-stage2}.
In Stage-2, the models with and without distillation performed comparably.
In Stage-3, it can be observed that the student model that has trained using knowledge distillation in the second stage (i.e., $S_3$) shows consistent improvement. This improvement is only slightly better than the model when distillation is not performed (i.e., $S_{3}^{'}$) in the second stage training.
This could be attributed to the fact that the second-pass mainly focuses on re-scoring the predicted first pass hypotheses instead of re-decoding them. Hence, its performance is limited by the first-pass decoder.

\begin{table}[t]
\caption{WERs(\%)/ SERs(\%) for teacher and smaller student models (denoted by $SS_2$ and $SS_3$) in Stage-2 and Stage-3.}
\centering
\resizebox{0.355\textwidth}{!}{%
\begin{tabular}{ c|c c|c c } 
 \toprule
 \multirow{2}{*}{\textbf{Model-ID}} & \multicolumn{2}{|c}{\textbf{test-clean}} & \multicolumn{2}{|c}{\textbf{test-other}} \\\cline{2-5}
      & \textbf{WER} & \textbf{SER} & \textbf{WER} & \textbf{SER} \\ 
 \midrule
 \midrule
 $T_{2}$ & 5.80 & 53.63 & 16.79 & 79.62 \\
 $SS_{2}$ & 6.25 & 55.99 & 17.99 & 81.69 \\
 $S_{2}$ & 6.24 & 55.95 & 17.77 & 81.35 \\
 \midrule
 $T_{3}$ & 5.87 & 53.55 & 16.92 & 80.16 \\
 $SS_{3}$ & 6.05 & 54.50 & 17.93 & 81.83 \\
 $S_{3}$ & 5.81 & 52.67 & 17.27 & 79.72 \\
 \bottomrule
\end{tabular}
}
\label{tab:ss}
\end{table}

\vspace{-0.2cm}
\subsection{Reducing parameters for two-pass student model}
\vspace{-0.1cm}
Since our student model $S_3$ performs slightly better than teacher model $T_3$, we also conduct an experiment where we reduce the model size of $S_3$ further from 36\%  to 55\% compared to the $T_3$ model. This is done by reducing the size of the LAS module of the student model in Stage-2 and Stage-3. These  student model with LAS size smaller than $T_2$ and $T_3$ are denoted by $SS_2$ and $SS_3$ respectively.

The results obtained with this experiment are given in Table~\ref{tab:ss}. 
It can be observed that the smaller student model $SS_3$ performs comparable to model $S_3$ and shows degradation of only 3\% (relative) while achieving a significant compression rate of 55\% compared to the teacher model $T_3$.
This is a reasonable compromise in the performance with an impressive gain achieved in the compression rate of 55\%, which is good enough for medium to high-end mobile devices.

\vspace{-0.2cm}
\section {Conclusion}
\vspace{-0.2cm}
In this paper, we propose a model compression technique for the two-pass streaming ASR model using knowledge distillation.
The proposed approach reduces the two-pass model size by 36\% while slightly improving the WER as compared to the teacher model.
Further, with the proposed approach, our smaller student model is able to achieve a high compression rate of 55\% compared to the two-pass teacher model while performing comparable to the teacher model in terms of WER. Needless to mention, the two-pass student model with the proposed compression approach has shown significantly better performance than that of the smaller two-pass model trained independently.  
We also perform detailed analyses of the proposed approach in different stages. 
In future, we plan to extend our approach  and perform a comparative study with the low-rank approximation methods.

\bibliographystyle{IEEEbib}
\bibliography{main}

\begin{thebibliography}{10}

\bibitem{he2019streaming}
Yanzhang He, Tara~N Sainath, Rohit Prabhavalkar, Ian McGraw, Raziel Alvarez,
  Ding Zhao, David Rybach, Anjuli Kannan, Yonghui Wu, Ruoming Pang, et~al.,
\newblock ``Streaming end-to-end speech recognition for mobile devices,''
\newblock in {\em IEEE International Conference on Acoustics, Speech and Signal
  Processing (ICASSP)}. IEEE, 2019, pp. 6381--6385.

\bibitem{sainath2019two}
Tara~N Sainath, Ruoming Pang, David Rybach, Yanzhang He, Rohit Prabhavalkar,
  Wei Li, Mirk{\'o} Visontai, Qiao Liang, Trevor Strohman, Yonghui Wu, et~al.,
\newblock ``Two-pass end-to-end speech recognition,''
\newblock {\em arXiv preprint arXiv:1908.10992}, 2019.

\bibitem{ganesh_ondevice}
Ganesh Venkatesh, Alagappan Valliappan, Jay Mahadeokar, Yuan Shangguan,
  Christian Fuegen, Michael~L. Seltzer, and Vikas Chandra,
\newblock ``Memory-efficient speech recognition on smart devices,''
\newblock in {\em IEEE International Conference on Acoustics, Speech and Signal
  Processing (ICASSP)}, 2021, pp. 8368--8372.

\bibitem{samsung_streaming}
Kwangyoun Kim, Kyungmin Lee, Dhananjaya Gowda, Junmo Park, Sungsoo Kim, Sichen
  Jin, Young-Yoon Lee, Jinsu Yeo, Daehyun Kim, Seokyeong Jung, et~al.,
\newblock ``Attention based on-device streaming speech recognition with large
  speech corpus,''
\newblock in {\em Proc. of ASRU}, 2019, pp. 956--963.

\bibitem{garg2019asru}
Abhinav Garg, Dhananjaya Gowda, Ankur Kumar, Kwangyoun Kim, Mehul Kumar, and
  Chanwoo Kim,
\newblock ``Improved multi-stage training of online attention-based
  encoder-decoder models,''
\newblock in {\em Proc. of ASRU}, 2019, pp. 70--77.

\bibitem{infra_paper}
Chanwoo Kim, Sungsoo Kim, Kwangyoun Kim, Mehul Kumar, Jiyeon Kim, Kyungmin Lee,
  Changwoo Han, Abhinav Garg, Eunhyang Kim, Minkyoo Shin, Shatrughan Singh,
  Larry Heck, and Dhananjaya Gowda,
\newblock ``End-to-end training of a large vocabulary end-to-end speech
  recognition system,''
\newblock in {\em { Proc. of ASRU }}, 2019, pp. 562--569.

\bibitem{sr_vd_is2020}
Abhinav Garg, Gowtham Vadisetti, Dhananjaya Gowda, Sichen Jin, Aditya
  Jayasimha, Youngho Han, Jiyeon Kim, Junmo Park, Kwangyoun Kim, Sooyeon Kim,
  Youngyoon Lee, Kyungbo Min, and Chanwoo Kim,
\newblock ``Streaming on-device end-to-end {ASR} system for privacy-sensitive
  voicetyping,''
\newblock in {\em Proc. of Interspeech}, 2020, pp. 3371--3375.

\bibitem{kim2020review_ondevice}
Chanwoo Kim, Dhananjaya Gowda, Dongsoo Lee, Jiyeon Kim, Ankur Kumar, Sungsoo
  Kim, Abhinav Garg, and Changwoo Han,
\newblock ``A review of on-device fully neural end-to-end automatic speech
  recognition algorithms,''
\newblock {\em arXiv preprint arXiv:2012.07974}, 2020.

\bibitem{graves2006connectionist}
Alex Graves, Santiago Fern{\'a}ndez, Faustino Gomez, and J{\"u}rgen
  Schmidhuber,
\newblock ``Connectionist temporal classification: labelling unsegmented
  sequence data with recurrent neural networks,''
\newblock in {\em Proceedings of the 23rd International Conference on Machine
  Learning}, 2006, pp. 369--376.

\bibitem{chan2015listen}
William Chan, Navdeep Jaitly, Quoc~V Le, and Oriol Vinyals,
\newblock ``Listen, attend and spell,''
\newblock {\em arXiv preprint arXiv:1508.01211}, 2015.

\bibitem{Gou_2021}
Jianping Gou, Baosheng Yu, Stephen~J. Maybank, and Dacheng Tao,
\newblock ``Knowledge distillation: A survey,''
\newblock {\em International Journal of Computer Vision}, vol. 129, no. 6, pp.
  1789–1819, Mar 2021.

\bibitem{chebotar2016distilling}
Yevgen Chebotar and Austin Waters,
\newblock ``Distilling knowledge from ensembles of neural networks for speech
  recognition.,''
\newblock in {\em Proc. of Interspeech}, 2016, pp. 3439--3443.

\bibitem{hinton2015distilling}
Geoffrey Hinton, Oriol Vinyals, and Jeff Dean,
\newblock ``Distilling the knowledge in a neural network,''
\newblock {\em arXiv preprint arXiv:1503.02531}, 2015.

\bibitem{fukuda2017efficient}
Takashi Fukuda, Masayuki Suzuki, Gakuto Kurata, Samuel Thomas, Jia Cui, and
  Bhuvana Ramabhadran,
\newblock ``Efficient knowledge distillation from an ensemble of teachers.,''
\newblock in {\em Proc. of Interspeech}, 2017, pp. 3697--3701.

\bibitem{albawi2017understanding}
Saad Albawi, Tareq~Abed Mohammed, and Saad Al-Zawi,
\newblock ``Understanding of a convolutional neural network,''
\newblock in {\em International Conference on Engineering and Technology
  (ICET)}. IEEE, 2017, pp. 1--6.

\bibitem{simonyan2014very}
Karen Simonyan and Andrew Zisserman,
\newblock ``Very deep convolutional networks for large-scale image
  recognition,''
\newblock {\em arXiv preprint arXiv:1409.1556}, 2014.

\bibitem{sak2014long}
Hasim Sak, Andrew~W Senior, and Fran{\c{c}}oise Beaufays,
\newblock ``Long short-term memory recurrent neural network architectures for
  large scale acoustic modeling,''
\newblock {\em arXiv preprint arXiv:1402.1128}, 2014.

\bibitem{sherstinsky2020fundamentals}
Alex Sherstinsky,
\newblock ``Fundamentals of recurrent neural network ({RNN}) and long
  short-term memory ({LSTM}) network,''
\newblock {\em Physica D: Nonlinear Phenomena}, vol. 404, pp. 132306, 2020.

\bibitem{kim2016sequence}
Yoon Kim and Alexander~M Rush,
\newblock ``Sequence-level knowledge distillation,''
\newblock {\em arXiv preprint arXiv:1606.07947}, 2016.

\bibitem{wong2016sequence}
Jeremy~HM Wong and Mark Gales,
\newblock ``Sequence student-teacher training of deep neural networks,''
\newblock in {\em Proc. of Interspeech}, 2016, pp. 2761--2765.

\bibitem{inaguma2021source}
Hirofumi Inaguma, Tatsuya Kawahara, and Shinji Watanabe,
\newblock ``Source and target bidirectional knowledge distillation for
  end-to-end speech translation,''
\newblock {\em arXiv preprint arXiv:2104.06457}, 2021.

\bibitem{kim2017improved}
Suyoun Kim, Michael~L Seltzer, Jinyu Li, and Rui Zhao,
\newblock ``Improved training for online end-to-end speech recognition
  systems,''
\newblock {\em arXiv preprint arXiv:1711.02212}, 2017.

\bibitem{takashima2018investigation}
Ryoichi Takashima, Sheng Li, and Hisashi Kawai,
\newblock ``An investigation of a knowledge distillation method for ctc
  acoustic models,''
\newblock in {\em IEEE International Conference on Acoustics, Speech and Signal
  Processing (ICASSP)}. IEEE, 2018, pp. 5809--5813.

\bibitem{kurata2020knowledge}
Gakuto Kurata and George Saon,
\newblock ``Knowledge distillation from offline to streaming {RNN-Transducer}
  for end-to-end speech recognition,''
\newblock {\em Proc. of Interspeech}, pp. 2117--2121, 2020.

\bibitem{panchapagesan2020efficient}
Sankaran Panchapagesan, Daniel~S Park, Chung-Cheng Chiu, Yuan Shangguan, Qiao
  Liang, and Alexander Gruenstein,
\newblock ``Efficient knowledge distillation for rnn-transducer models,''
\newblock {\em arXiv preprint arXiv:2011.06110}, 2020.

\bibitem{las}
William Chan, Navdeep Jaitly, Quoc Le, and Oriol Vinyals,
\newblock ``Listen, attend and spell: A neural network for large vocabulary
  conversational speech recognition,''
\newblock in {\em IEEE International Conference on Acoustics, Speech and Signal
  Processing (ICASSP)}, 2016, pp. 4960--4964.

\bibitem{graves2012sequence}
Alex Graves,
\newblock ``Sequence transduction with recurrent neural networks,''
\newblock {\em arXiv preprint arXiv:1211.3711}, 2012.

\bibitem{panayotov2015librispeech}
Vassil Panayotov, Guoguo Chen, Daniel Povey, and Sanjeev Khudanpur,
\newblock ``Librispeech: an asr corpus based on public domain audio books,''
\newblock in {\em IEEE International Conference on Acoustics, Speech and Signal
  processing (ICASSP)}. IEEE, 2015, pp. 5206--5210.

\bibitem{sennrich2015neural}
Rico Sennrich, Barry Haddow, and Alexandra Birch,
\newblock ``Neural machine translation of rare words with subword units,''
\newblock {\em arXiv preprint arXiv:1508.07909}, 2015.

\bibitem{wanli2013research}
Zhang Wanli and Li~Guoxin,
\newblock ``The research of feature extraction based on mfcc for speaker
  recognition,''
\newblock in {\em Proceedings International Conference on Computer Science and
  Network Technology}. IEEE, 2013, pp. 1074--1077.

\bibitem{park2019specaugment}
Daniel~S Park, William Chan, Yu~Zhang, Chung-Cheng Chiu, Barret Zoph, Ekin~D
  Cubuk, and Quoc~V Le,
\newblock ``Specaugment: A simple data augmentation method for automatic speech
  recognition,''
\newblock {\em arXiv preprint arXiv:1904.08779}, 2019.

\bibitem{srivastava2014dropout}
Nitish Srivastava, Geoffrey Hinton, Alex Krizhevsky, Ilya Sutskever, and Ruslan
  Salakhutdinov,
\newblock ``Dropout: a simple way to prevent neural networks from
  overfitting,''
\newblock {\em The Journal of Machine Learning Research}, vol. 15, no. 1, pp.
  1929--1958, 2014.

\bibitem{zeyer2018improved}
Albert Zeyer, Kazuki Irie, Ralf Schl{\"u}ter, and Hermann Ney,
\newblock ``Improved training of end-to-end attention models for speech
  recognition,''
\newblock {\em arXiv preprint arXiv:1805.03294}, 2018.

\bibitem{kingma2014adam}
Diederik~P Kingma and Jimmy Ba,
\newblock ``Adam: A method for stochastic optimization,''
\newblock {\em arXiv preprint arXiv:1412.6980}, 2014.

\bibitem{vaswani2017attention}
Ashish Vaswani, Noam Shazeer, Niki Parmar, Jakob Uszkoreit, Llion Jones,
  Aidan~N Gomez, Lukasz Kaiser, and Illia Polosukhin,
\newblock ``Attention is all you need,''
\newblock {\em arXiv preprint arXiv:1706.03762}, 2017.

\end{thebibliography}
\end{document}